\documentclass[12pt]{article}
\usepackage[left= 2.5cm, right= 2.5cm, bottom=2.5cm, top=2.5cm]{geometry}
\usepackage{amsmath}
\usepackage{amssymb}
\usepackage{times}
\usepackage{amsfonts}
\usepackage{graphicx}
\usepackage{colordvi}
\usepackage[english]{babel}
\usepackage{verbatim}
\usepackage{color}
\usepackage{mathrsfs}
\usepackage{epsfig}
\usepackage{wrapfig}
\usepackage{floatflt}
\usepackage{enumitem}
\usepackage{setspace}
\usepackage[font=small,format=plain,justification=justified]{caption}
\title{Magnetic resonance imaging of spin-wave transport and interference in a magnetic insulator} 
\vspace{12pt}

\author
{Iacopo Bertelli,$^{1,2}$ Joris J. Carmiggelt,$^{1}$ Tao Yu,$^{1}$ Brecht G. Simon,$^{1}$\\ Coosje C. Pothoven,$^{1}$ Gerrit E. W. Bauer,$^{1,3}$ Yaroslav M. Blanter,$^{1}$\\ Jan Aarts$^{2}$ $\&$ Toeno van der Sar$^{1,*}$}
\begin{document} 
\baselineskip24pt
\maketitle
\noindent$^{1}$ Department of Quantum Nanoscience, Kavli Institute of Nanoscience, Delft University of Technology, Lorentzweg 1, 2628 CJ, Delft, The Netherlands\\
$^{2}$ Huygens - Kamerlingh Onnes Laboratorium, Leiden University, Niels Bohrweg 2, 2300 RA, Leiden, The Netherlands\\
$^{3}$ Institute for Materials Research \& WPI-AIMR \& CSRN, Tohoku University, Sendai 980-8577, Japan\\
$^*$ Corresponding author. Email:  T.vanderSar@tudelft.nl.
\onehalfspacing
\subsection*{Abstract}
Spin waves - the elementary excitations of magnetic materials - are prime candidate signal carriers for low-dissipation information processing. Being able to image coherent spin-wave transport is crucial for developing interference-based spin-wave devices. We introduce a platform for probing coherent spin waves based on magnetic resonance imaging with electron spins in diamond. Focusing on a thin-film magnetic insulator, we quantify spin-wave amplitudes, visualize the dispersion, and demonstrate time-domain measurements of spin-wave packets. We use our platform to study spin-wave interference, revealing uni-directional, autofocused spin-wave patterns with frequency-controlled numerical apertures. A theoretical analysis explains the patterns in terms of chiral spin-wave excitation and stray-field coupling to the sensor spins. These results pave the way for probing spin waves in atomically thin magnets, even when embedded between opaque materials.
\linespread{1}
\newpage
\singlespace
\subsection*{Introduction}
Over the last few decades, the desire to understand and control spin transport, and to use it in information technology, has invigorated the field of spintronics. A central goal of the field is to provide information processing based on the spin of the electron instead of its charge, and thereby avoid the heating associated with charge currents. As heating is currently the main obstacle for increasing computational speed, spin-based information processing may provide the next transformative change in information technology.\\
\indent Promising signal carriers for low-dissipation information transport are spin waves \cite{Kruglyak Magnonics, Chumak Magnon Spintronics} - the collective spin excitations of magnetic materials. Spin waves exist even in electrically insulating magnets, where they are able to propagate inherently free of the dissipative motion of charge. They can have nanometer wavelengths and gigahertz frequencies well suited for chip-scale device technologies and interference-based spin-wave logic circuits \cite{Chumak Magnon Spintronics}. Consequently, a growing research field focuses on spin-wave devices such as interconnects, interferometers, transistors, amplifiers, and spin-torque oscillators \cite{Hille SW multiplexer, Hille Magnon Transistor, Cornelissen, Demidov STO, Hille SW DirCoupler}.\\
\indent Being able to image coherent spin waves in thin-film magnets is crucial for developing spin-wave device technology. In recent years, several advanced techniques for imaging coherent spin waves have been developed, such as time-resolved scanning transmission x-ray microscopy (TR-STXM) \cite{STXM}, Brillouin light scattering (BLS) \cite{BLS}, and Kerr microscopy \cite{Kerr}, which rely on a spin-dependent optical response of a magnetic material. Here we introduce a new method, based on phase-sensitive magnetic resonance imaging of the microwave magnetic fields generated by a spin wave using electronic sensor spins in a diamond chip (Fig. 1A). This approach combines an 
ability to detect low magnetic fields with a sensitivity to their chirality, making it well suited for spin-wave imaging in magnetic thin films. We use this approach to study spin-wave transport in a thin-film magnetic insulator, revealing uni-directional excitation of spin waves that autofocus, interfere, and produce magnetic stray fields with a chirality that matches that of the natural precession of the sensor spins. We perform a theoretical analysis of the chiral spin-wave excitation and stray-field coupling to the sensor spins and show that it accurately describes the observed spatial spin-wave maps.\\
\indent We detect the magnetic fields generated by spin waves using electron spins associated with nitrogen-vacancy (NV) lattice defects in diamond \cite{Rondin review}. These spins can be initialized and read out optically, and manipulated with high fidelity via microwave excitation. Over the last decade, NV magnetometry has emerged as a powerful platform for probing static and dynamic magnetic phenomena in condensed matter systems \cite{Toeno review}. Key is an NV-sample distance tunable between 10-1000 nm that is well matched with the length scales of spin textures such as magnetic domain walls, cycloids, vortices, and skyrmions \cite{Gross BFO, Rondin vortex Nat Comm, Dovzhenko skyrmion}, as well as those of dynamic phenomena such as spin waves \cite{Toeno SWs, Toeno chem pot, Andrich SWs, Kikuchi SWs, Wolfe SWs}. Recent experiments demonstrated that NV magnetometry has the sensitivity required for imaging the static magnetization of monolayer van-der-Waals magnets \cite{Thiel CrI3}. Here, we develop NV-based magnetic resonance imaging into a platform for studying coherent spin waves via the gigahertz magnetic fields they generate.
\subsection*{Results}
Our imaging platform consists of a diamond chip hosting a dense layer of shallowly implanted NV spins. We position this chip onto a thin film of yttrium iron garnet (YIG) - a ferrimagnetic insulator with record-high magnetic quality \cite{Serga YIG magnonics} (Fig. 1A-B). The typical distance between the diamond and the magnetic film is $\sim1~\mu$m (Supplementary Material). We excite spin waves using microwave striplines microfabricated onto the YIG. When the spin-wave frequency matches an NV electron spin resonance (ESR) frequency, the oscillating magnetic stray field $\mathit{B}_{SW}$ drives NV spin transitions \cite{Toeno SWs, Andrich SWs} that we detect through the NV’s spin-dependent photoluminescence (Methods). By tuning the external static magnetic field $\mathit{B}_0$, we sweep the NV ESR frequencies through the spin-wave band, thereby probing spin waves with different wavelength (Fig. 1C).\\
\indent We start by characterizing the NV photoluminescence as a function of $\mathit{B}_0$ and the frequency $\omega_{MW}$ of a microwave drive current sent through the stripline, at $\sim5~\mu m$ distance from the stripline edge (Fig. 1D). This microwave current generates an oscillating magnetic field that drives ESR transitions of the NV spins directly, but also excites spin waves in the YIG film that can drive NV ESR transitions via their magnetic stray field (Fig. 1A). The dips in the observed NV photoluminescence correspond to the ESR frequencies of the NV spins in the diamond (Fig. 1D, Methods). Importantly, we observe an enhanced contrast for the $\omega_-$ transition when $B<B_0^{(2)}$. In this region, the excited spin waves efficiently drive the $\omega_-$ ESR transition.\\
\indent We image the spin waves excited by the stripline in the YIG film by characterizing the contrast of the $\omega_-$ ESR transition as a function of the distance to the stripline (Fig. 2A). We do so by tuning the magnetic field such that the $\omega_-$ frequency is a few 100 MHz above the bottom of the spin-wave band, thereby exciting spin waves in the film. To gain the phase sensitivity required for detecting the individual wave fronts of these propagating spin waves, we let their stray field interfere with an additional, externally applied microwave magnetic field $B_{REF}$ that is spatially homogeneous and has the same frequency (Methods). As formulated mathematically below, this interference leads to a spatial standing-wave pattern in the total magnetic field that drives the NV ESR transition with a spatial periodicity equal to the spin-wave wavelength. We can thus rapidly visualize the spin waves by measuring the ratio between the NV photoluminescence with and without applied microwaves (Fig. 2A).\\
\indent Quantifying the amplitude of a spin wave is a challenging task for any technique since the coupling between spin wave and probe is often not well known. With NV magnetometry, however, we accurately measure the microwave magnetic field generated by a spin wave as described by Maxwell’s equations. We can therefore determine the amplitude of a spin wave of known direction and ellipticity with high confidence by solving a well-defined inverse problem.\\
\indent To illustrate the concept, we formulate the magnetic stray field of a spin wave traveling perpendicularly to the static magnetization (such as the one in Fig. 2B) in the reference frame depicted in Fig. 1A with transverse magnetization
\begin{equation}
\mathbf{m_\perp}(y)=m_\perp^0 \Re[e^{i(k_y y-\omega t)}  (\mathbf{\hat{y}} -i\eta \mathbf{\hat{x}})]
\label{eqn1}
\end{equation}
where ky, $\omega$ and η are the wavenumber, angular frequency and ellipticity of the spin wave, respectively, t is the time, and hats denote unit vectors. This spin wave produces a magnetic stray field above the film that rotates in the xy plane (see Supplementary Material and \cite{Yu PRL})
\begin{equation}
\mathbf{B}_{SW} (y)=-B_{SW}^0 \Re[e^{i(k_y y-\omega t)} (\mathbf{\hat{y}} +i\mathrm{sgn}(k_y) \mathbf{\hat{x}})]
\label{eqn2}
\end{equation}		
where $B_{SW}^0=\mu_0 m_\perp^0 (1+\mathrm{sgn}(k_y )\eta)|\mathbf{k}|de^{-|k_y |x_0}/2$, $x_0$  is the NV-YIG distance, and $d$ is the thickness of the YIG film.\\
\indent The handedness of $\mathbf{B}_{SW}$ is opposite to that of $\mathbf{m}_\perp$ for a spin wave traveling to the right (i.e., with $k_y>0$, as in Fig. 2B), which drives the $\omega_-$  (rather than the $\omega_+$) NV spin transition (Supplementary Material). Moreover, the amplitude $B_{SW}^0$ depends on the propagation direction and degree of ellipticity $\eta$ of the spin wave: those traveling to the right (left) generate a stronger field above (below) the magnetic film. Therefore, only the $\omega_-$ transition of NV centers to the right of the stripline in Fig. 2B is excited (Supplementary Material). The resulting NV spin rotation rate (Rabi frequency) $\omega_{Rabi}$ is determined by the interference between the spin-wave field and the reference field $B_{REF}$:
\begin{equation}
\omega_{Rabi} (y)=\sqrt{2} \gamma|B_{SW}^0  \cos^2⁡(\frac{\phi}{2}) e^{ik_y y}-B_{REF}|
\label{eqn3}
\end{equation}
where $\phi = 35^\circ$ is the angle w.r.t. the film of the NV centers used in Fig. 2 and $\gamma/2\pi=28$ GHz/T is the (modulus of the) electron gyromagnetic ratio. Fitting the data in Fig. 2B by Eq. \ref{eqn3} (including a spatial decay, see Supplementary Material), we extract a spin-wave amplitude $m_\perp^0=0.039(4)M_S$ at the location of the stripline and a decay length of $1.15(9)$ mm, corresponding to a Gilbert damping parameter $1.3(2)\cdot 10^{-4}$.\\
\indent By tuning the externally applied magnetic field we sweep the NV ESR frequency through the spin-wave band and access spin waves with different wavelengths (Fig. 3A), as schematically described in Fig. 1C. In Fig. 3A-B we visualize the individual spin-wave fronts using the interference between the direct stripline field and the stray field of the propagating spin wave. We extract the spin-wave dispersion from the frequency-dependence of the wavelength (Fig. 3C). This dispersion matches the one calculated using values of the saturation magnetization $M_s$ and film-thickness $d$ determined by independent measurements (Supplementary Material).\\
\indent Traveling spin-wave packets can be used for pulsed quantum control of distant spins such as those of the NV centers \cite{Andrich SWs, Kikuchi SWs}. Understanding the distance-dependent response of the spins to an applied control sequence requires knowledge of the spin-wave group velocity. We demonstrate a time-domain characterization of the spin-wave propagation using pulsed control of the NV spins (Fig. 3D-E). In our measurement scheme (Fig. 3D) a spin-wave pulse excited by the stripline only flips the NV spins into the $m_s=-1$ state at a location under study if this pulse arrives before a reference pulse (applied via a wire located above the sample). Measurements as a function of time between spin wave and reference pulses and distance from the stripline reveal the spin-wave packet in the time domain and allow the extraction of the group velocity (Fig. 3E). We find a velocity of $3.6(2)$ km/s at a frequency of $2.169$ GHz and wavelength $12~\mu$m, consistent with the the YIG spin-wave dispersion (Supplementary Material).\\
\indent The 2 mm-long stripline used in Fig. 2 and Fig. 3 corresponds to an effectively one-dimensional situation. We now turn to spin waves injected by a shorter stripline with a length comparable to the scanned area (Fig. 4A). We observe a focused emission pattern that is dominated by spin-wave beams traveling at specific angles (Fig. 4B-C). Such "caustics" occur when the dispersion is strongly anisotropic \cite{Hille caustics, Wolfe caustics}, and can be understood in terms of stationary points in the iso-frequency curves in reciprocal space (Fig. 4D). In optics, such an iso-frequency curve $k_z=k_z (k_y)$ is called "slowness" curve, since it is perpendicular to the group velocity $\mathbf{v_G}=\mathbf{\nabla}_k \omega(k)$. The states for which the angle of the group velocity $\theta=-\arctan⁡(dk_z (k_y)/dk_y )$ is stationary along the curve, i.e. when $d\theta/dk_y \propto d^2 k_z (k_y)/dk_y^2 =0$, dominate emission, generating high-intensity spin-wave beams. The external magnetic field and the drive frequency can tune the beam direction and intensity, providing opportunities to optimize the efficiency of spin wave-mediated driving of distant spins at target locations (Supplementary Material).\\
\indent Finally, we image the interference between spin waves excited by two adjacent striplines on the YIG chip (Fig. 4E-F), which highlights the unidirectional spin-wave excitation by the striplines and shows rich interference patterns radiating from the three crossing points of the main caustics (i.e. $\sim80~\mu$m from the striplines edge). The strongly anisotropic spin-wave dispersion causes a triangular "dark" region between the striplines in which no spin waves are detected, because spin waves traveling at small angles with respect to the equilibrium magnetization direction or having large wavenumbers are neither efficiently excited (Supplementary Material) nor efficiently detected due to the ~1 micron NV-sample distance. The observed directionality and interference patterns agree well with linear response calculations of the non-local dynamic susceptibility and the spatial profile of the microwave drive field, as described in the Supplementary Material. These quantitative measurements of the spin-wave-generated chiral magnetic stray fields illustrate the power of NV-based magnetic resonance imaging in magnonics.
\subsection*{Discussion}
Our results demonstrate that ensembles of NV spins in diamonds enable quantitative, phase-sensitive magnetic imaging of coherent spin waves in thin-film magnets. By magnetic imaging we can see through optically opaque materials, opening the way to study e.g. top-gated magnetic films and multilayer systems. The typical NV-magnet distances in this work are $0.5-2~\mu$m (limited by e.g. dust particles), comparable to the diffraction-limited spatial resolution of NV-ensemble-based magnetic imaging, enabling phase-sensitive observation of spin waves with wavelengths on this scale. Shallow NV centers in diamond chips that are wafer-bonded to (i.e., in direct contact with) a magnetic sample of interest will allow the detection of spin waves with wavelengths comparable to the implantation depth of the NV centers of a few nanometers \cite{Rosskopf NV depth}. Phase-sensitive imaging of spin waves with wavelengths below the diffraction limit could be enabled using specialized NV control sequences such as phase encoding schemes \cite{Arai sub-diff imaging}. Furthermore, the techniques presented here are directly transferrable to single-NV scanning probe microscopes with real-space resolution on the 10 nm scale \cite{Balasub scanning}.\\
\indent Our results pave the way for studying spin waves in other magnetic material systems such as magnetic nanodevices and atomically thin magnets. NV magnetometry works at cryogenic temperatures \cite{Thiel cryo scanning, Pelliccione scanning, Wrachtrup cryo scanning}, allowing studies of magnets with low Curie temperatures such as complex-oxide or van-der-Waals magnets. Because the dipole density per unit area $M_sd =3.6\cdot10^3 \mathrm{\mu_B/nm^2}$ of the YIG film studied here is only about 2 orders of magnitude above the $16~\mathrm{\mu_B/nm^2}$ of the monolayer van der Waals magnet $\mathrm{CrI_3}$ \cite{Thiel CrI3}, the magnetic stray fields generated by spin waves in such monolayer magnets are within the sensitivity range of NV-based magnetic imaging.

\subsection*{Acknowledgments}
\textbf{Funding:} This work was supported by the Dutch Research Council (NWO) as part of the Frontiers of Nanoscience (NanoFront) program, through the Projectruimte grant 680.91.115, JSPS KAKENHI Grant No. 19H006450, and by the Kavli Institute of Nanoscience Delft. \textbf{Author contributions:} I.B., J.J.C., and T.v.d.S. designed the experiment. I.B. fabricated the diamond-YIG samples, realized the imaging setup, performed the NV measurements and analyzed the data. B.G.S. prepared the diamonds. C.C.P. performed the VNA measurements, for which J.J.C. fabricated the samples. T.Y., Y.B., and G.B. developed the theoretical model describing spin-wave caustics and interference. I.B. and T.v.d.S. wrote the manuscript with help from all co-authors. \textbf{Competing interests:} The authors declare no competing financial interest. \textbf{Data and materials availability:} All data contained in the figures will be made available at Zenodo.org upon publication. Additional data related to this paper may be requested from the authors.
\newpage
\begin{figure}[htbp]
\begin{center}\vspace{-32pt}

{\includegraphics[width=0.9\textwidth]{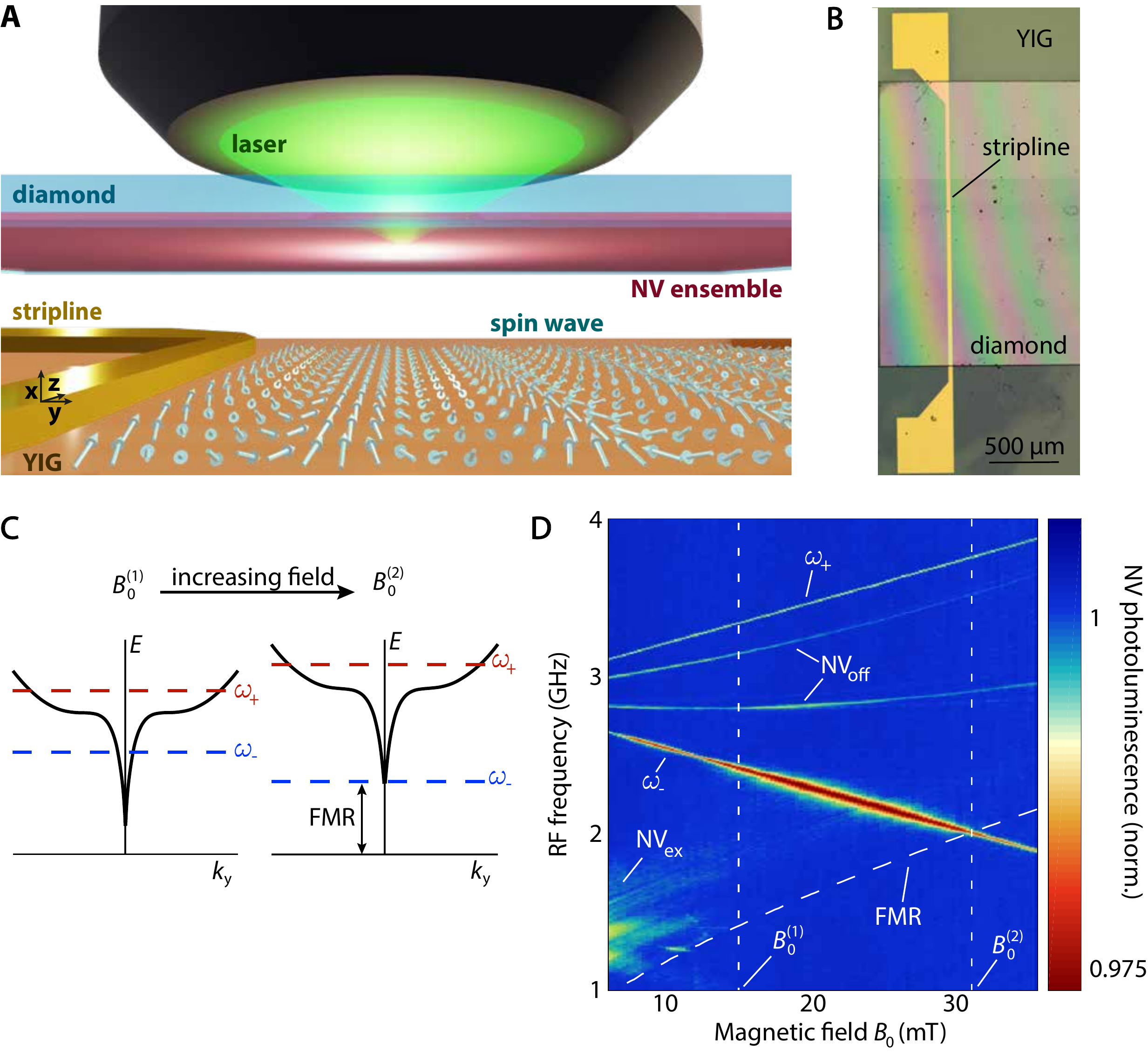}}
\end{center}\vspace{-12pt}
\caption{\textbf{Fig. 1}. \textbf{Imaging spin waves using nitrogen-vacancy (NV) spins in diamond as magnetic field sensors.} \textbf{(A)} Idea of the experiment. A diamond hosting a dense, near-surface layer of NV spins is placed on top of a film of yttrium iron garnet (YIG). The NV spins detect the magnetic stray fields of spin waves excited by a microwave stripline. The NVs were implanted 20 nm below the surface of the $\sim40~\mu$m thick diamond chip. The YIG was grown on a gadolinium gallium garnet (GGG) substrate (YIG thickness 245 nm, GGG thickness 500 $\mu$m).
\textbf{(B)} Optical micrograph of an NV-center-containing diamond chip on top of a YIG substrate equipped with a gold stripline. The static field $B_0$ is applied along the stripline and at $\phi =35^\circ$ relative to the sample plane to align it with one of the four possible crystallographic orientations of NV centers in diamond. The YIG magnetization points along the stripline and is in-plane to within a few degrees for the small magnetic fields $B_0<35$ mT used in this work (Damon-Eshbach configuration).
\textbf{(C)} Sketch showing how the NV ESR frequencies $\omega_-$ and $\omega_+$ are swept over the Damon-Eshbach branch of the spin-wave dispersion (black line) by tuning $B_0$. At $B_0=B_0^{(2)}$ (right panel) $\omega_-$ is resonant with the $k=0$ ferromagnetic resonance (FMR).
\textbf{(D)} Normalized NV photoluminescence as a function of $B_0$ and the frequency of a microwave drive current sent through an on-chip stripline, measured at $\sim5~\mu$μm from the stripline edge on a sample with a $\sim2.5~\mu$m-wide stripline. Indicated are the ESR transitions of the NV centers that are aligned with $B_0$ in the electronic ground ($\omega_\pm$) and excited ($\mathrm{NV_{ex}}$) state, and the ESR transitions of the off-axis NV families ($\mathrm{NV_{off}}$). $B_0^{(1)}$ and $B_0^{(2)}$ correspond to the fields indicated in \textbf{(C)}.
The ferromagnetic resonance frequency (FMR) is calculated from the independently determined saturation magnetization (Supplementary Material), matching the frequency above which the contrast of the $\omega_-$ transition is enhanced.}
\end{figure}
\newpage
\begin{figure}[htbp]
\begin{center}
{\includegraphics[width=1\textwidth]{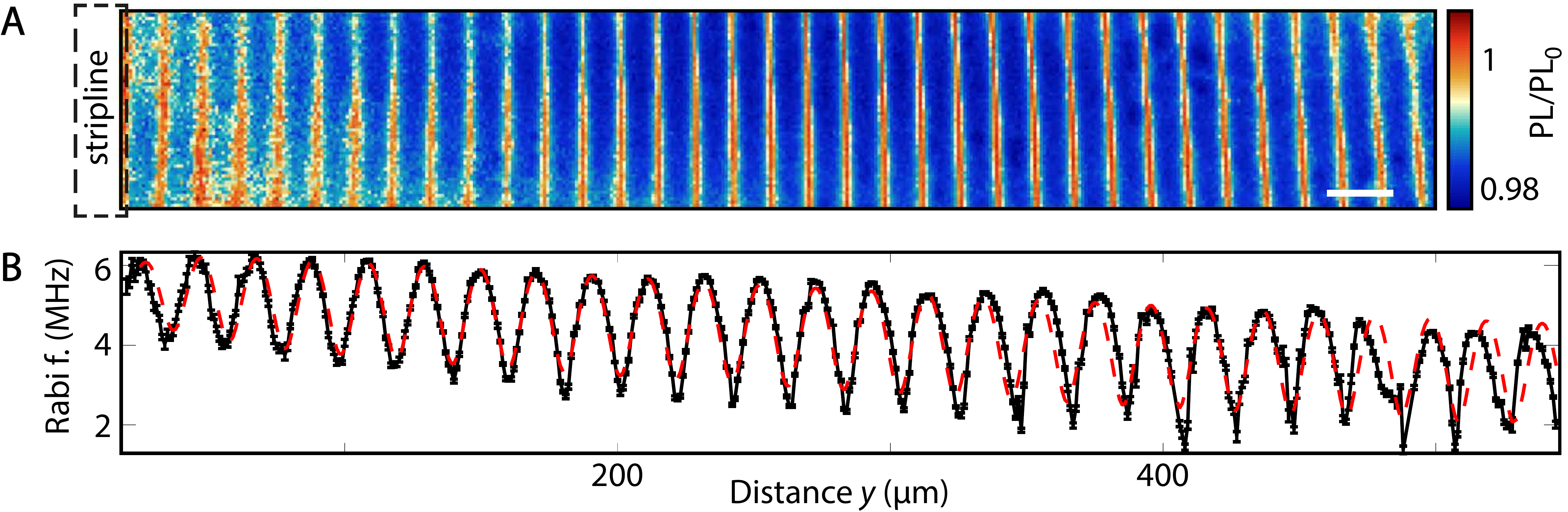}}
\end{center}
\caption{\textbf{Fig. 2}. \textbf{Magnetic imaging of coherent spin waves. } \textbf{(A)} Spatial map of the NV ESR contrast at $B_0$ = 25 mT when a spin wave of frequency $\omega_{SW}=\omega_-=2\pi\cdot2.17$ GHz is excited by a microwave current through the stripline (length 2 mm, width 30 $\mu$m, thickness 200 nm) located at the left edge of the image. For each pixel, the NV photoluminescence (PL) with applied microwaves is normalized to the NV photoluminescence ($\mathrm{PL_0}$) without applied microwaves. The NV-YIG distance at the stripline was 1.8(2) $\mu$m, as determined by fitting the measured field produced by a DC current in the stripline (Supplementary Material). Scale bar: 20 $\mu$m. \textbf{(B)} NV spin rotation rate (Rabi frequency $\omega_{Rabi}/2\pi$) as a function of the distance from the stripline. $\omega_{SW}=\omega_-=2\pi\cdot2.11$ GHz and $B_0 = 27$ mT. In both \textbf{(A)} and \textbf{(B)}, the microwave current is split between the stripline and a bonding wire located at $\sim100~\mu$m above the YIG, oriented along the spin-wave propagation direction to generate a spatially homogeneous magnetic field, creating a standing-wave interference pattern (see text).}
\end{figure}
\newpage
\begin{figure}[th]
\begin{center}\vspace{-13pt}
{\includegraphics[width=1\textwidth]{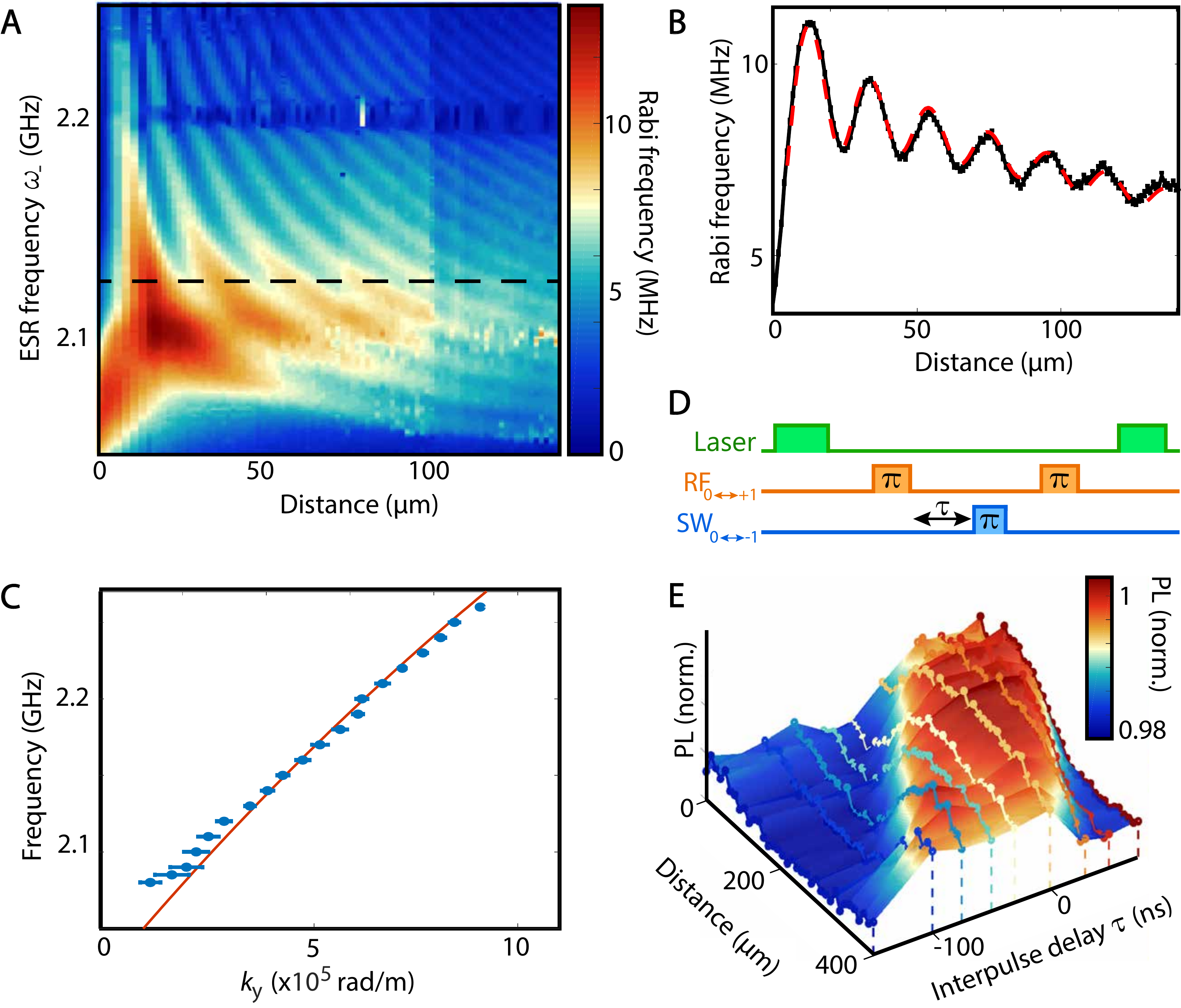}}
\end{center}
\caption{\textbf{Fig. 3. Spin-wave dispersion in the space- and time-domains.} \textbf{(A)} Spatial map of the NV spin rotation rate (Rabi frequency) as a function of the microwave drive frequency and distance from the stripline. The horizontal feature at 2.2 GHz matches the calculated frequency of the first perpendicular spin-wave mode (see Supplementary Material). \textbf{(B)} Horizontal linecut of \textbf{(A)} with fit (red line) at 2.119 GHz. \textbf{(C)} Blue dots: measured spin-wave frequency vs. wave number (perpendicular to the stripline) extracted from the measurement in \textbf{(A)}. Red line: calculated spin-wave dispersion. \textbf{(D)} Pulse sequence used to study spin-wave packets in the time domain (data shown in \textbf{(E)}): the NV spins are prepared in $m_s=0$ using a 1 $\mu$s green-laser pulse. Two $\pi$-pulses from an external microwave source, separated by 100 ns, excite the $0\leftrightarrow+1$  transition. After a delay time $\tau$ from the end of the first pulse, a spin-wave-mediated $\pi$-pulse acts on the $0\leftrightarrow-1$  transition. Finally, the NV spin is read-out by measuring the NV photoluminescence during the first 400 ns of a green laser pulse. \textbf{(E)} Normalized NV photoluminescence (PL) as a function of distance from the spin-wave injector and inter-pulse delay τ. Negative times indicate a spin-wave packet generated before the first RF pulse. For example, for an inter-pulse delay of τ=-100 ns the signal rises at 360 $\mu$m, indicating that spin waves travel 360 $\mu$m in 100 ns, i.e. with a group velocity of 3.6 km/s. Circles represent data, the colored surface is an interpolation.}
\end{figure}
\newpage
\begin{figure}[th]
\begin{center}
\includegraphics[width=1\textwidth]{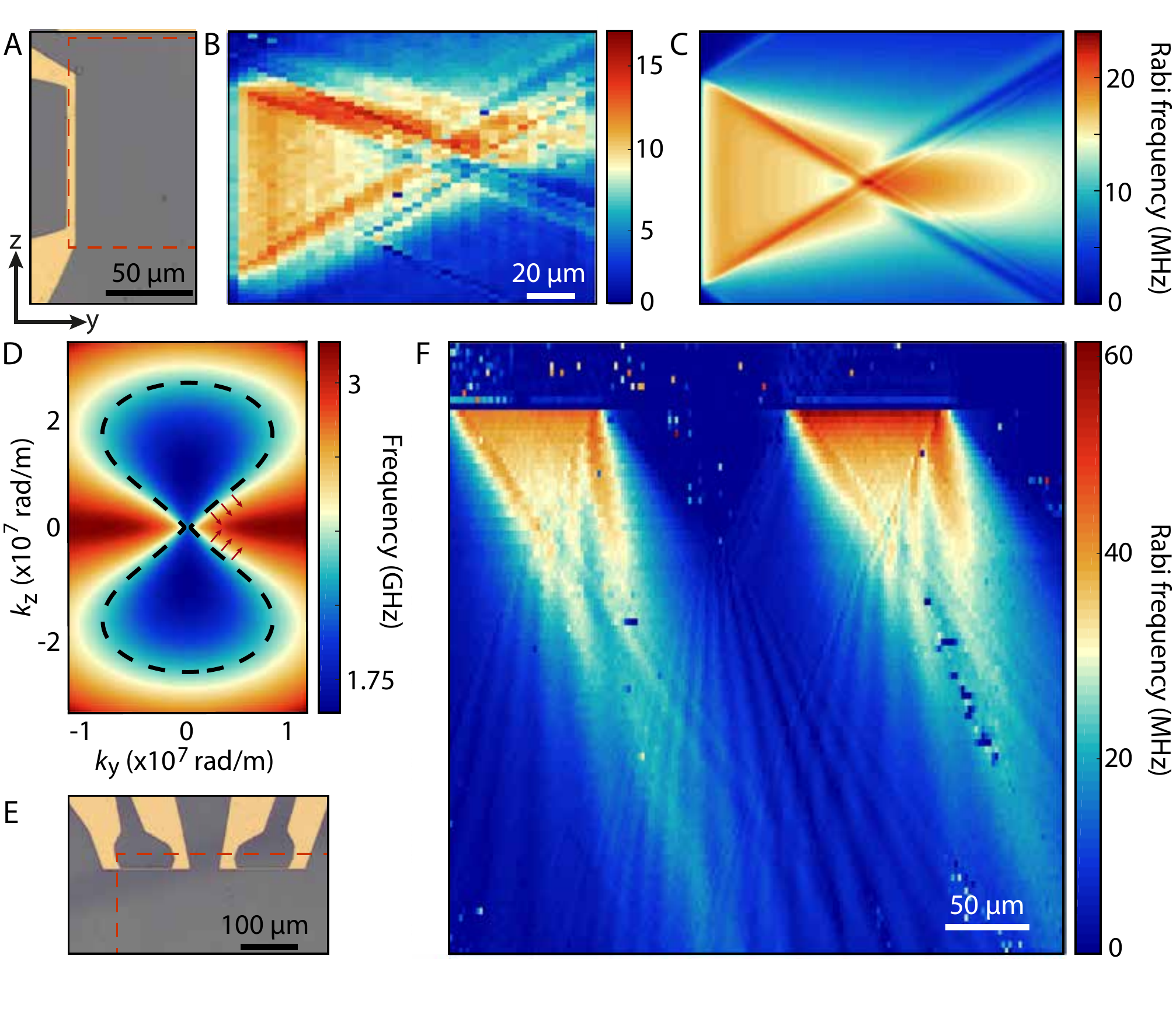}
\end{center}
\caption{\textbf{Fig. 4. Imaging interference and caustics of spin waves excited by one and two short striplines.} \textbf{(A)} Optical micrograph of the stripline (width 5 $\mu$m) used to excite spin waves. The dashed red lines indicate the region where \textbf{(B)} is acquired. \textbf{(B)} Rabi frequency map relative to the dashed region of \textbf{(A)} for $B_0= 27.1$ mT and $\omega/2\pi=2.11$ GHz. The small asymmetry is attributed to an independently measured $\sim100~\mu$T gradient in $B_0$ over the imaged area caused by the finite size of the permanent magnet used to generate the static field. \textbf{(C)} Simulation of the emission pattern observed in \textbf{(B)}. \textbf{(D)} Calculated two-dimensional spin-wave dispersion relation $\omega(k_y,k_z)/2π$ at $B_0= 20.5$ mT. The dashed line is an iso-frequency contour at 2.292 GHz, indicating which wavevectors can be excited at this frequency and field. Red arrows indicate the direction of the spin-wave caustics. \textbf{(E)} Optical micrographs of the two injector striplines of width 2.5 $\mu$m. The dashed lines indicates the region where \textbf{(F)} is acquired. \textbf{(F)} Rabi frequency map under simultaneous driving of the two striplines, showing uni-directional excitation of autofocused spin-wave patterns that interfere and drive NV Rabi oscillations via their chiral magnetic stray fields.}
\label{Rabi_frequency}%
\end{figure}
\end{document}